%% file: attractorDistribution.tex
\documentclass[review, 3p, twocolumn]{elsarticle}

\journal{Computers and Electronics in Agriculture}

\biboptions{authoryear}

\usepackage{amsmath}
\usepackage{amssymb}

\usepackage{booktabs}

\usepackage{tikz}
\usetikzlibrary{decorations.pathreplacing, positioning, shapes, arrows.meta}
\tikzset{>=latex}

\usepackage{subcaption}

\usepackage[noend]{algpseudocode}
\usepackage{algorithm}

\usepackage{subcaption}

\usepackage{layouts}

\usepackage{url}

\usepackage{enumitem}

\usepackage{hyperref}
\hypersetup
{
    pdftitle = {Enforcing Mean Reversion in State Space Models for Prawn Pond Water Quality Forecasting},
    pdfauthor = {Joel Janek Dabrowski},
}

\makeatletter
\newcommand\thefontsize[1]{{#1 The specified font size is: \f@size pt\par}}
\makeatother

\newcommand{\titleText}{Enforcing Mean Reversion in State Space Models for Prawn Pond Water Quality Forecasting}

\begin{document}
	
	\begin{frontmatter}
		
		\title{\titleText}

		\author[addr1]{Joel Janek Dabrowski \corref{cor1}}
		\ead{joel.dabrowski@data61.csiro.au}
		
		\author[addr2]{Ashfaqur Rahman}
		\ead{ashfaqur.rahman@data61.csiro.au}
		
		\author[addr3]{Daniel Edward Pagendam}
		\ead{dan.pagendam@data61.csiro.au}
		
		\author[addr3]{Andrew George}
		\ead{andrew.george@data61.csiro.au}
		
		\address[addr1]{Data61, CSIRO, St Lucia, QLD, Australia}
		\address[addr2]{Data61, CSIRO, Sandy Bay, TAS, Australia}
		\address[addr3]{Data61, CSIRO, Dutton Park, QLD, Australia}
		
		\cortext[cor1] {Corresponding author; St Lucia, QLD, 4067,
			Australia.}
		
		\begin{abstract}
			The contribution of this study is a novel approach to introduce mean reversion in multi-step-ahead forecasts of state-space models. This approach is demonstrated in a prawn pond water quality forecasting application. The mean reversion \textit{constrains} forecasts by gradually drawing them to an average of previously observed dynamics. This corrects deviations in forecasts caused by irregularities such as chaotic, non-linear, and stochastic trends. The key features of the approach include (1) it enforces mean reversion, (2) it provides a means to model both short and long-term dynamics, (3) it is able to apply mean reversion to select structural state-space components, and (4) it is simple to implement. Our mean reversion approach is demonstrated on various state-space models and compared with several time-series models on a prawn pond water quality dataset. Results show that mean reversion reduces long-term forecast errors by over 60\% to produce the most accurate models in the comparison.
		\end{abstract}
		
		
	\end{frontmatter}

	
	
	\input{introduction}
	\input{relatedWork}
	\input{methods}
	\input{meanReversion}

	\input{datasets}
	\input{models}
	\input{results}

	\input{comparisonResults}
	\input{conclusion}

	\bibliography{Bibliography}
	
\end{document}

%% file: introduction.tex

\section{Introduction}

In aquaculture prawn farming, managing water quality is key for maximising quantity, quality, and health of the stock. For example, high levels of prawn mortality can occur due to anoxia and hypoxia if dissolved oxygen (DO) drop to extreme values \citep{robertson2006Australian}. By forecasting important water quality variables, farmers are provided with the tools to take preemptive measures that encourage favourable pond conditions.

Long-term forecasting can be a challenging task with complex environmental processes such as prawn ponds. In this study, we take advantage of the fact that many natural processes exhibit some form of mean reversion. This is commonly found where the process seeks a state of equilibrium. For example, the long-term trend (a week or more) of pond water temperature typically varies within some bounds. These bounds are maintained as the underlying process seeks thermodynamic equilibrium within a changing environment. Without knowledge of the underlying process, the longer-term dynamics can appear as a slowly varying stochastic trend. 

Forecasting such processes can be challenging when stochastic trends cause forecasts to deviate. Models should realistically incorporate some form of constraint or bounds. Our hypothesis is that such a constraint can be imposed by modelling the stochastic variations with a fixed \textit{attractor distribution} that long-term trends are drawn towards. In this form, the long-term behaviour of the process may have some stable, marginal distribution when integrated over time (long periods of time or just the recent past).

In this study we propose a novel approach to introduce an attractor distribution in non-stationary state-space models. The attractor distribution models previously observed dynamics. Mean reversion is enforced through introducing pseudo-observations into the Kalman filter during forecasting. These pseudo-observations are samples of the attractor distribution mean. The result is that the filtering operation during forecasting naturally draws the forecasts towards the mean of the previously observed dynamics.

The proposed approach can model both short and long-term dynamics and it allows for the selection of which state space components should be mean reverting. Furthermore, the approach is easily implemented using the standard Kalman filter and it has broad appeal as it addresses problems that are found in many domains other than aquaculture.

Our contributions are: (1) we provide an approach to enforcing mean reversion in state-space models (to our knowledge, no other studies have introduced any form of mean reversion into state space models for constraining forecasts), (2) we demonstrate this approach on several state-space models in a real-world aquaculture application, and (3) we compare our approach with several time series models.

This paper is organised as follows: In section \ref{sec:relatedWork}, we review related forecasting literature. Section \ref{sec:ldsAndFiltering}, provides an overview of the linear dynamic system (LDS) and the Kalman filter with the purpose of introducing our mean reversion approach described in section \ref{sec:attractorDistribution}. The aquaculture problem and datasets used in this study are presented in section \ref{sec:datasets}. In section \ref{sec:models} we demonstrate how our approach is applied to state space forecasting models and results are provided in section \ref{sec:results}. In section \ref{sec:comparison} a comparison of our approach with several forecasting methods is provided. The study is concluded in section \ref{sec:conclusion}.

%% file: relatedWork.tex

\section{Related Work}
\label{sec:relatedWork}
    
\subsection{Forecasting Models}

Many industries and disciplines rely multi-step-ahead forecasting. A wide range of forecasting methods exist in the literature \citep{Gooijer200625}. Statistical models include state-space models, regression models, exponential smoothing, Box-Jenkins models (such as the autoregressive moving average (ARMA) model), long memory models, autoregressive conditional heteroscedastic (ARCH), and generalised ARCH (GARCH) models. Nonlinear machine learning models have also been extensively explored for forecasting. Neural networks in particular have a relatively large body of literature \citep{zhang2005Neural, Zhang1998Forecasting, Ruiz2018Energy}.

State-space models are generative, probabilistic, interpretable, and flexible \citep{durbin2012time}. As generative models, they are able to handle missing data and forecasting functionality is inherent. As probabilistic models, they provide a natural representation of uncertainty in a forecast. State-space models are interpretable as they are designed based on structural analysis of a problem and naturally incorporate explanatory variables. This is in contrast with data driven models such as neural networks and ARMA models, which are considered as black-box models.

\subsection{Multi-Step-Ahead Forecasting}

Multi-step-ahead forecasting is a challenging task as it requires a complete model of the short and longer-term dynamics. Short-term modelling is required to model the dynamics \textit{between} the forecast time-steps. Longer-term modelling is required to model the dynamics \textit{across} the several time-step forecasts.

The general approach to long-term forecasting is to model the long-term trend of the time series and ignore short term dynamics. Such models can be obtained using time series analysis methods such as regression models, state-space models, Box-Jenkins models, and recurrent neural networks \citep{Kandil2001Overview, Soman2010review, Granger2007Long}. It is however possible to combine long and short-term forecasts as discussed in the review presented by \citet{Andrawis2011Combination}. The authors note that there seems to be little work in the literature relating to such combinations, despite their effectiveness. 

The approach we present in this study does not require combining long and short term-models. Rather, it provides a means to naturally include both short-term and long-term dynamics in a single model. The short-term dynamics are modelled directly in the state-space model. The long-term dynamics are modelled using mean reversion and the attractor distribution.

\subsection{Mean Reversion}

Many phenomena should realistically be modelled with some form of limiting distribution for long-term forecasts. For example, interest rates are often modelled through the use of mean-reverting stochastic processes, such as the Ornstein-Uhlenbeck process (e.g. the Vasicek model \citep{Vasicek1977equilibrium} or the CIR model \citep{Cox1985Theory}). The dynamics are limited to Brownian motion with a tendency towards the origin \citep{Pavliotis2014Stochastic}. Though Brownian motion is not stationary, a linear damping term in the Ornstein-Uhlenbeck process can cause the process to become stationary. The generalised Ornstein-Uhlenbeck process is a natural continuous time analogue of the AR(1) process with random i.i.d. components \citep{rao2012time}.

The ARMA model also exhibits mean reversion, but the moving-average allows for mean-reversion to occur more gradually. In general, AR and ARMA models are limited to modelling only stationary sequences \cite{box2015time}. Non-stationary components such as trend and seasonality are removed from the time series through differencing such as in the Autoregressive Integrated Moving Average (ARIMA) model.

The ARMA and ARIMA models may be framed as state-space models \citep{durbin2012time}. In general, state-space models are not limited to stationary series and provide expressive power through latent variables. State-space models are however not necessarily mean reverting. Our proposed approach provides the means to enforce mean reversion in state-space models.

\subsection{Water Quality Modelling}

In water quality modelling applications, several ecosystem-based models have been proposed for variables such as DO \citep{Ginot1994Estimating, lu1996stochastic, madsen2007modelling, Xu2016Deterministic}. These are complex multivariable models that require precisely determined parameters pertaining to biological and physical processes. Various data-driven approaches have also been used for modelling and forecasting water quality variables. These include neural networks \citep{Zhang2019Applying, Ta2018Research, Ren2018method, Dabrowski2018Prediction, Fernandez2016Soft, Schmid2006Artificial, dogan2009Modeling, Rankovic2010Neural, Basant2010Linear, He2011Abiotic, Ahmed2014Prediction} and other machine learning models \citep{shi2019Prediction,xu2017Prediction, Olyaie2017Comparative, Duan2016Water}.

\citet{dabrowski2018State} describe two data-driven state-space models for modelling DO, pH, and temperature in prawn ponds. These models provide a compromise between ecosystem models and machine learning models. They are data-driven unlike ecosystem models, and are not black-box models like many machine learning models. The proposed mean reversion approach is tested on these models in the context of forecasting water quality variables.

%% file: methods.tex

\section{The Linear Dynamic System and Filtering}
\label{sec:ldsAndFiltering}

\subsection{The Linear Dynamic System}

\begin{figure}[t]
	\centering
	\input{figures/fig_latentDynamicModel}
	\caption{Graphical model representation of the latent dynamic model such as the linear dynamic system.}
	\label{fig:latentDynamicModel}
\end{figure}
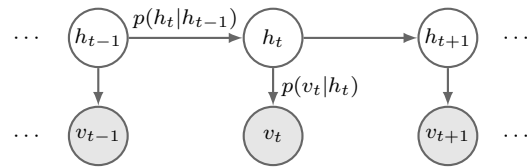

The linear dynamic system (LDS) is a state-space model that assumes linear-Gaussian dynamics \citep{barber2012bayesian, thrun2005probabilistic, murphy2012machine}. Consider a system comprising a latent or hidden variable $h_t$ that evolves over time, $t = 1,\dots,T$. The system provides an observable variable $v_t$ from which measurements can be made. The observable variable is considered to have been emitted from the latent variable $h_t$. Assuming a first order Markov process, the graphical model describing this system is illustrated in \figurename{~\ref{fig:latentDynamicModel}}. The edges between the latent variables describe the transition distribution $p(h_t | h_{t-1})$. The edges between the latent and observable variables describe the emission distribution $p(v_t | h_t)$.

Linear-Gaussian assumptions in the LDS result in the following state-space equations \citep{petris2009dynamic, grewal2015kalman}
\begin{align}
&h_t = A h_{t-1} + \eta_t^h \label{eq_lds_ht}\\
&v_t = B h_{t} + \eta_t^v \label{eq_lds_vt}
\end{align}
The variable $h_t$ is the state vector, $A$ is the state transition matrix, and $\eta_t^h \sim \mathcal{N}(0, \Sigma^h)$ is the state noise vector (where $\Sigma$ denotes a covariance matrix). The variable $v_t$ is the observation vector, $B$ is the emission or measurement matrix, and $\eta_t^v \sim \mathcal{N}(0, \Sigma^v)$ is the measurement noise vector. In continuous time, state-space equations are given by \citep{grewal2015kalman, zarchan2000fundamentals, durbin2012time}
\begin{align}
&\dot{h}(t) = \breve{A} h(t) + \eta^h(t) \label{eq_lds_ht2}\\
&v(t) = \breve{B} h(t) + \eta^v(t) \label{eq_lds_vt2}
\end{align}
where $\breve{A}$ and $\breve{B}$ denote the continuous time state and emission matrices.

\subsection{The Kalman Filter (KF)}
\label{sec:theKalmanFilter}

Inference in the LDS involves calculating $p(h_t | v_{1:t})$, which is the probability distribution over the current latent variable given all past observations \citep{barber2012bayesian, murphy2012machine}. The linear-Gaussian assumption allows for a closed-form inference algorithm known as the Kalman filter (KF) \citep{kalman1960new}. The filtered distribution is represented as a Gaussian with mean $f_t$ and covariance $F_t$. The KF algorithm recursively repeats a prediction and update step. In the prediction step, the Gaussian distributions $p(h_t | v_{1:t-1})$ and $p(v_t | v_{1:t-1})$ are computed. The mean and covariance relating to $p(h_t | v_{1:t-1})$ distributions are given by
\begin{align}
\mu^{h}_t &= A f_{t-1} \label{eq_kf_muh}\\
\Sigma^{hh}_t &= A F_{t-1} A^T + \Sigma^h \label{eq_kf_Shh}
\end{align}
The mean and covariance relating to $p(v_t | v_{1:t-1})$ are given by
\begin{align}
\mu^{v}_t & = B \mu^h_t \label{eq_kf_muv}\\
\Sigma^{vv}_t &= B \Sigma^{hh}_t B^T + \Sigma^v  \label{eq_kf_Svv}
\end{align}
Additionally, the cross-covariance between the latent and observed variables is given by
\begin{align}
\Sigma^{hv}_t &= \Sigma^{hh}_t B^T  \label{eq_kf_Svh}
\end{align}

The predictions are updated with the latest observations to provide the parameters for the filtered distribution. These parameters are given by
\begin{align}
f_t &= \mu^h_t + K_t (v_t - \mu^v_t) \label{eq_kf_ft} \\
F_t &= (I - K_t B)\Sigma^{hh}_t \label{eq_kf_Ft}
\end{align}
where $I$ is the identity matrix and $K_t$ is the Kalman gain given by
\begin{align}
K_t &= \Sigma^{hv}_t (\Sigma^{vv}_t)^{-1} \label{eq_kf_K1}\\
&= (\Sigma^{hh}_t B^T) (B \Sigma^{hh}_t B^T + \Sigma^v)^{-1} \label{eq_kf_K2}
\end{align}

\subsection{Forecasting with the LDS}

The filtered distribution is computed at each time using equations (\ref{eq_kf_ft}) and (\ref{eq_kf_Ft}) with observations $v_t$. During forecasting, the prediction equations (\ref{eq_kf_muh}), (\ref{eq_kf_Shh}), (\ref{eq_kf_muv}), and (\ref{eq_kf_Svv}) are used with no observations. For multiple forecasts into the future, $f_{t-1}$ and $F_{t-1}$ in equations (\ref{eq_kf_muh}) and (\ref{eq_kf_Shh}) can be replaced with $\mu_{t-1}$ and $\Sigma^{hh}_{t-1}$ respectively. Multiple forecasts are thus generated by sequentially sampling from the model.

Any forecasts made for times $t+i$, $i>0$ are calculated based on the dynamics of the model at time $t$. These dynamics are contained in the filtered distribution at time $t$. If the filtered distribution at time $t$ is not representative of the long-term trend, long-term forecasts may be inaccurate.

\subsection{Nonlinear and Non-Gaussian Filtering}

The Kalman filter is a closed form solution for a linear-Gaussian model. If a system is nonlinear or non-Gaussian, approximate filtering methods such as the extended Kalman filter (EKF), the unscented Kalman filter (UKF) \citep{Julier1997New}, or Monte Carlo methods such as the particle filter \citep{Gordon1993Novel} and ensemble Kalman filter (enKF) \citep{Evensen1994Sequential} are required. In this study the EKF is used. The EKF approximates a nonlinear function by linearising around the current state mean estimate \citep{zarchan2000fundamentals}.

%% file: figures/fig_latentDynamicModel.tex
%

\def\horisep{2.3}
\def\vertsep{-1.3}

\begin{tikzpicture}
	\footnotesize
	\tikzstyle{every path}=[->,draw=black!60, thick]
	\tikzstyle{vNode}=[circle,draw=black!60, fill=black!10,minimum size=22pt,inner sep=0pt] 
	\tikzstyle{hNode}=[circle,draw=black!60, minimum size=22pt,inner sep=0pt] 
	\tikzstyle{label}=[];
	\node[label] (dots) at (-0.4*\horisep,0) {$\cdots$};
	\node[hNode] (htm1) at (0*\horisep,0) {$h_{t-1}$};
	\node[hNode] (ht) at (1*\horisep,0) {$h_{t}$};
	\node[hNode] (htp1) at (2*\horisep,0) {$h_{t+1}$};
	\node[label] (dots) at (2.4*\horisep,0) {$\cdots$};
	\node[label] (dots) at (-0.4*\horisep,\vertsep) {$\cdots$};
	\node[vNode] (vtm1) at (0*\horisep,\vertsep) {$v_{t-1}$};
	\node[vNode] (vt) at (1*\horisep,\vertsep) {$v_{t}$};
	\node[vNode] (vtp1) at (2*\horisep,\vertsep) {$v_{t+1}$};
	\node[label] (dots) at (2.4*\horisep,\vertsep) {$\cdots$};
	
	\draw[->] (htm1) -- (ht) node[midway, above] {$p(h_t | h_{t-1})$};
	\draw[->] (ht) -- (htp1) node[midway, above] {};
	\draw[->] (htm1) -- (vtm1) node[midway, above] {};
	\draw[->] (ht) -- (vt) node[midway, right] {$p(v_t | h_t)$};
	\draw[->] (htp1) -- (vtp1) node[midway, above] {};
\end{tikzpicture}

%% file: meanReversion.tex

\section{Mean Reversion and the Attractor Distribution}
\label{sec:attractorDistribution}

\subsection{Forecast Deviation In State-Space Models}
\label{sec:illustration}

State-space time series models are comprised of several distinct components such as trend, seasonal, and noise (disturbances) \citep{durbin2012time, commandeur2007introduction, west2013bayesian, hyndman2008forecasting, harvey1990forecasting, petris2009dynamic}. The trend component is often represented in the form of a polynomial model. Especially models such as the first-order-polynomial Dynamic Linear Model (DLM) perform well for relatively short-term forecasting but can fail in longer term forecasts \citep{west2013bayesian}.
Irregularities such as slowly varying stochastic trends can shift the forecast trajectory off course. Mean reversion corrects the deviant forecast by drawing it back towards the attractor distribution mean.

\subsection{Attractor Distribution and the Central Limit}

The proposed approach is to use an attractor distribution to draw the forecasts to the mean of a distribution that approximates the central limit. \citet{Spall1984Asymptotic} proved the central limit theorem for the Kalman filter under certain conditions. These conditions include the standard Kalman filter assumptions as well as uniform complete observability and controllability. The intention of the study was to investigate the asymptotic nature of the Kalman filter. \citet{Aliev1999Evaluation} furthered this study by investigating the convergence rate of the central limit theorem for the Kalman filter.

To approximate the mean of the central limit distribution, the average over all filtered posterior distributions (see Section \ref{sec:theKalmanFilter}) is computed up to time $t$. That is
\begin{align}
\label{eq_mr}
f_{\infty} \approx \frac{1}{t} \sum_{i=1}^{t} f_i.
\end{align}
This approximation is used as the mean of the attractor distribution.

It is also possible to compute a weighted average where more emphasis is given to recent dynamics. A geometric progression can be used to obtain an exponential weighted average as follows\footnote{Note that the form $f_{\infty} \approx \lambda \sum_{i=1}^{t} f_i (1-\lambda)^{t-i}$ can be used if $\lambda$ and $t$ are chosen such that $\lambda \sum_{i=1}^{t} (1-\lambda)^{t-i} \approx 1$.}
\begin{align}
\label{eq_wmr}
f_{\infty} \approx \frac{\sum_{i=1}^{t} f_i (1-\lambda)^{t-i}}{\sum_{i=1}^{t}(1-\lambda)^{t-i}},
\end{align}
where $\lambda$ is some constant in the range $0 < \lambda \leq 1$. This provides a form of exponential smoothing \citep{brown1959statistical, holt1957forecasting, Winters1960Forecasting} in the mean reversion.

\subsection{Mean Reversion Through Filtering}

To draw the forecast to the attractor distribution mean, it is proposed that the forecasts be filtered with the attractor distribution as an observable variable. That is, set $v_t = f_\infty$ as a pseudo-observation during forecasting. The filtered distribution can be written as \citep{thrun2005probabilistic}
\begin{align}
\label{eq_bayes}
p(h_{t} | v_{1:t}) \propto p(v_t|h_t) p(h_t | v_{1:t-1})
\end{align}
The first term can be viewed as a likelihood of the observation given the model state. The second term can be viewed as a prior describing the predicted model state given previous observations. By using the attractor distribution as the observable variable, the likelihood describes the probability of the attractor distribution given the current model state. If this likelihood is low, it implies a mismatch between what the model is forecasting and what is expected asymptotically.

To understand how filtering draws the forecast to the attractor distribution, consider the Kalman filter update equation (\ref{eq_kf_ft}). The filtered mean is the current prediction $\mu^h_t$, that is updated with a weighted difference between observation $v_t$ and the prediction $\mu^v_t$. The weighting factor for the error is the Kalman gain. Equation (\ref{eq_kf_ft}) provides a mechanism to correct the model prediction with an observable variable $v_t$. If $v_t$ is the attractor distribution, the forecast will be corrected according to the attractor distribution.

\subsection{Parameters}

To define the emission matrix $B$ for the attractor distribution pseudo-observations, consider that $B$ provides a mapping from the space of $h_t$ to the space of $v_t$. The matrix $B$ can be manipulated to map only certain components from the latent variable space. Non-zero values can be placed in $B$ corresponding to components which should be mean reverting in nature. For example, non-zero values could be placed in $B$ corresponding to trend components that should exhibit mean reversion behaviour. Zeros can be placed in $B$ corresponding to components which should not be mean reverting in nature. For example, seasonal components may be left to oscillate throughout a forecast. A demonstration of this is presented in Section \ref{sec:models}.

To define the measurement noise covariance $\Sigma^v$ for the attractor distribution pseudo-observations, consider that $\Sigma^v$ represents a form of uncertainty of the observation. By adjusting the uncertainty, the rate of convergence of the forecast to the attractor distribution mean can be manipulated. The Kalman gain defines the level of correction. Consider the representation of the Kalman gain in (\ref{eq_kf_K2}). The expression comprises $B$, $\Sigma^{hh}_t$, and $\Sigma^v$. $B$ is defined as discussed above and $\Sigma^{hh}_t$ is computed from the prediction. With these defined, the Kalman gain can thus be adjusted by manipulating $\Sigma^v$. If $\Sigma^v$ is set to zeros, indicating the extreme level of certainty of $v_t$, the Kalman gain reduces as follows
\begin{align}
K_t &= (\Sigma^{hh}_t B^T) (B \Sigma^{hh}_t B^T + \mathbf{0})^{-1} \nonumber\\
&= (\Sigma^{hh}_t B^T) (B^{-T} (\Sigma^{hh}_t)^{-1} B^{-1}) \nonumber\\
&= B^{-1}
\end{align}
If $K_t = B^{-1}$, the filtered mean in (\ref{eq_kf_ft}) reduces to $f_t = v_t$, which is the attractor distribution mean. If $\Sigma^v$ is set to infinite values along its diagonal to indicate an extreme level of uncertainty of $v_t$, (\ref{eq_kf_ft}) reduces to $f_t = \mu^h_t$, which is mean proposed by the model. That is, with infinite values in $\Sigma^v$, the attractor distribution will be ignored. 

By manipulating the uncertainty represented by $\Sigma^v$, the level of correction of the forecasts is controlled. This correction is performed over multiple steps during filtering. The result is that the rate of convergence of a forecast to the attractor distribution mean is determined by $\Sigma^v$.

%% file: datasets.tex

\section{Datasets}
\label{sec:datasets}

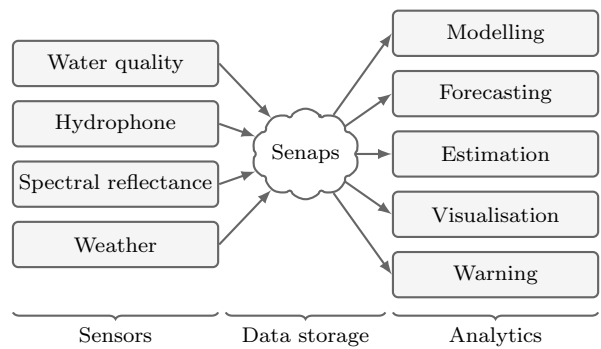
\begin{figure}[t]
	\centering
	\input{figures/fig_decisionSupportSystem}
	\caption{Aquaculture prawn farm decision support system.}
	\label{fig:decisionSupportSystem}
\end{figure}

This study fits within a broader context of a system that is being developed for aquaculture prawn farms. Several sensors have been deployed into prawn ponds for monitoring water quality related parameters.
These sensors include water quality sensors, hydrophones, spectral reflectance, and weather sensors. The sensor data is uploaded to a central cloud-based system (Senaps). Several decision support tasks are performed on the stored data. The framework of the decision support system is illustrated in \figurename{~\ref{fig:decisionSupportSystem}}. In this study, the modelling and forecasting of dissolved oxygen (DO), pH, and temperature in prawn ponds are considered. The mean reversion approach described in this study is applied to data collected within this decision support system.

The dataset used in this study comprises of DO, pH, and temperature readings taken from two prawn ponds. The first pond is a large 0.18ha grow-out pond and the second pond is a small 0.022ha nursery pond. The samples are taken at 15 minute intervals over a period of 88 days.

The datasets variables are seasonal in nature. Many water quality variables such as DO, pH and temperature follow diurnal fluctuations \citep{boyd1998pond}. Carbon dioxide ($\text{CO}_2$) is continually produced in the pond through respiration by organisms such as prawn and plankton. During the day, plant-based organisms use solar radiation for photosynthesis. Through photosynthesis, $\text{CO}_2$ is absorbed and oxygen is released. Thus, DO increases and $\text{CO}_2$ decreases during the day. At night photosynthesis ceases. The result is that DO decreases and $\text{CO}_2$ increases at night. $\text{CO}_2$ reacts with water to form carbonic acid. Increased acidity reduces the pH levels in the pond. Fluctuating $\text{CO}_2$ thus causes fluctuating pH. Furthermore, water temperature naturally fluctuates with the changes in solar radiation over a 24-hour period.

Water quality variables may also vary in an aperiodic manner \citep{boyd1998pond}. Irregular variations may be caused by weather-related variations and biological activity such as algal blooms. Such variations can produce the slow varying irregular or nonlinear fluctuations that cause forecast deviations.

%% file: figures/fig_decisionSupportSystem.tex

\def\horisep{2.50cm}
\def\vertsep{0.8cm}

\begin{tikzpicture}
	\tikzstyle{every path}=[->,draw=black!60, thick]
	\tikzstyle{sNode}=[draw=black!60, fill=black!4,  shape=rectangle, minimum width=2.7cm, minimum height=0.6cm, inner sep=0pt, rounded corners=2pt] 
	\tikzstyle{cNode}=[draw=black!60, fill=black!0,  shape=cloud, aspect=4, minimum height=1.2cm, inner sep=0pt, rounded corners=2pt] 
	\tikzstyle{annot} = [text width=2.4cm, text centered]
	\tikzstyle{every node} = [font=\footnotesize]
	
	\path (0*\horisep,3.5*\vertsep) node[sNode] (sens1) {Water quality};
	\path (0*\horisep,2.5*\vertsep) node[sNode] (sens2) {Hydrophone};
	\path (0*\horisep,1.5*\vertsep) node[sNode] (sens3) {Spectral reflectance};
	\path (0*\horisep,0.5*\vertsep) node[sNode] (sens4) {Weather};
	\path (1*\horisep,2*\vertsep) node[cNode] (cloud) {Senaps};
	\path (2*\horisep,4*\vertsep) node[sNode] (out1) {Modelling};
	\path (2*\horisep,3*\vertsep) node[sNode] (out2) {Forecasting};
	\path (2*\horisep,2*\vertsep) node[sNode] (out3) {Estimation};
	\path (2*\horisep,1*\vertsep) node[sNode] (out4) {Visualisation};
	\path (2*\horisep,0*\vertsep) node[sNode] (out5) {Warning};
	\draw[->] (sens1.east) -- (cloud);
	\draw[->] (sens2.east) -- (cloud);
	\draw[->] (sens3.east) -- (cloud);
	\draw[->] (sens4.east) -- (cloud);
	\draw[->] (cloud) -- (out1.west);
	\draw[->] (cloud) -- (out2.west);
	\draw[->] (cloud) -- (out3.west);
	\draw[->] (cloud) -- (out4.west);
	\draw[->] (cloud) -- (out5.west);
	
	\pgfmathsetlengthmacro\loca{0*\horisep-1.35cm}
	\pgfmathsetlengthmacro\locb{0*\horisep+1.35cm}
	\draw[-,annot,decoration={brace,mirror,raise=0pt},decorate] (\loca,-0.5) -- node[below=2pt] 
	{Sensors} (\locb,-0.5);
	
	\pgfmathsetlengthmacro\loca{1*\horisep-1.05cm}
	\pgfmathsetlengthmacro\locb{1*\horisep+1.05cm}
	\draw[-,annot,decoration={brace,mirror,raise=0pt},decorate] (\loca,-0.5) -- node[below=2pt] 
	{Data storage} (\locb,-0.5);
	
	\pgfmathsetlengthmacro\loca{2*\horisep-1.35cm}
	\pgfmathsetlengthmacro\locb{2*\horisep+1.35cm}
	\draw[-,annot,decoration={brace,mirror,raise=0pt},decorate] (\loca,-0.5) -- node[below=2pt] 
	{Analytics} (\locb,-0.5);
\end{tikzpicture}

%% file: models.tex

\section{Applied State-Space Models}
\label{sec:models}

\citet{dabrowski2018State} presented two models for modelling water quality parameters in prawn ponds. The first model is a LDS with a local linear trend component (constant velocity process) and a seasonal component. The second model is a nonlinear model that provides a means to model the seasonal amplitude using a local linear trend component. The UKF was used for inference in this non-linear model. These models will be used in this study, however the EKF algorithm will be used instead of the UKF algorithm. The intention is to improve the long-term (a week or more) forecasting capability of these models using the proposed mean reversion approach.

\subsection{Linear Model}
\label{sec:linearModel}

The observations of the linear model are modelled with a seasonal, trend and noise component as follows
\begin{align}
\label{eq_linearOut}
v_t = \alpha_t \sin(\omega t) + \gamma_t + \eta^v_t
\end{align}
The seasonal component $\alpha_t \sin(\omega t)$ is modelled with a sinusoid with amplitude $\alpha_t$, the trend $\gamma_t$ is modelled with as a continuous local linear trend model, and the noise $\eta^v_t$ is white Gaussian noise. The  Let $\psi_t = \alpha_t \sin(\omega t)$ such that \citep{dabrowski2018State}
\begin{align*}
h(t) =
\begin{bmatrix}
\gamma_t &
\dot{\gamma}_t &
\psi_t &
\dot{\psi}_t
\end{bmatrix}^T
\end{align*}
The state transition matrix in continuous time, denoted by $\breve{A}$ is then given by
\begin{align*}
\breve{A} = \begin{bmatrix}
0 & 1 & 0 & 0 \\
0 & 0 & 0 & 0 \\
0 & 0 & 0 & 1 \\
0 & 0 & \omega^2 & 0
\end{bmatrix}
\end{align*}
This matrix is converted to discrete time using a Laplace transform or the Taylor series expansion \citep{zarchan2000fundamentals}
\begin{align}
\label{eq:taylorSeries}
A =e^{\breve{A} \Delta t} = I + \breve{A} \Delta t + \frac{(\breve{A} \Delta t)^2}{2!} + \frac{(\breve{A} \Delta t)^3}{3!} + \cdots
\end{align}
where $\Delta t$ is the sample rate.

The emission matrix maps the elements from the latent variable space to the observed variable space according to (\ref{eq_linearOut}). The emission matrix is thus given by
\begin{align*}
B = \begin{bmatrix}
1 & 0 & 1 & 0 \\
\end{bmatrix}
\end{align*}

The attractor distribution is defined to draw the forecasts to a fixed mean of previously observed dynamics. For the linear model, mean reversion is applied to the trend component. Thus, the attractor distribution is defined to approximate the central limit of $\gamma_t$. The following emission matrix for the attractor distribution can thus be used
\begin{align*}
B =
\begin{bmatrix}
1 & 0 & 0 & 0
\end{bmatrix}.
\end{align*}
In this form, mean reversion is only enforced on $\gamma_t$ and not on the seasonal component $\psi_t$.

With the attractor distribution having a single dimension, the variance $\Sigma^v_t$ is a real number. The value is manually set to provide reasonable uncertainty bounds and to match the mean reversion settling time with the slowly varying irregular component of the data. As discussed in Section \ref{sec:attractorDistribution}, smaller values provide quicker settling times and narrower uncertainty bounds. Larger values provide slower settling times and wider uncertainty bounds. Suitable values can generally be found with a brief search over the sequence $10^i, ~ i \in \mathbb{Z}$ and further refined if necessary. A search can also be conducted using repeated random subsampling validation approaches.

\subsection{Nonlinear Model}

The linear model is independent of the sinusoidal amplitude $\alpha_t$ in (\ref{eq_linearOut}) \citep{dabrowski2018State}. Including the amplitude as a component in the state-space representation results in a nonlinear model. The amplitude is modelled as a latent variable with a constant velocity process such that
\begin{align*}
h(t) =
\begin{bmatrix}
\gamma_t &
\dot{\gamma}_t &
\alpha_t &
\dot{\alpha}_t &
\sin(\omega t) &
\cos(\omega t)
\end{bmatrix}^T
\end{align*}
The state transition matrix in continuous time is given by
\begin{align*}
\breve{A} =
\begin{bmatrix}
0 & 1 & 0 & 0 & 0 & 0 \\
0 & 0 & 0 & 0 & 0 & 0 \\
0 & 0 & 0 & 1 & 0 & 0 \\
0 & 0 & 0 & 0 & 0 & 0 \\
0 & 0 & 0 & 0 & 0 & 1 \\
0 & 0 & 0 & 0 & -\omega^2 & 0
\end{bmatrix}
\end{align*}
This matrix is converted to discrete time using (\ref{eq:taylorSeries})

The trend element is added to a product of the amplitude and sinusoidal elements as indicated in (\ref{eq_linearOut}). This results in a nonlinear emission model. Let $b(h_{t}) = \alpha_t \sin(\omega t) + \gamma_t$ such that
\begin{align*}
v_t = b(h_{t}) + \eta_t^v
\end{align*}
The EKF approach is to approximate the nonlinear function $b(h_{t})$ as a linearisation around the current state estimate. This linear approximation is the tangent to $b(h_{t})$ at the current state estimate. Thus, the emission matrix is given by \citep{zarchan2000fundamentals}
\begin{align*}
B = \left.  \frac{\partial b(h)}{\partial h} \right|_{h = f_t}
\end{align*}
That is, $B$ is given by the Jacobian
\begin{align*}
B &=
\begin{bmatrix}
\frac{\partial b(h)}{\partial \gamma_t}
& \frac{\partial b(h)}{\partial \dot{\gamma}_t}
& \frac{\partial b(h)}{\partial \alpha_t}
& \frac{\partial b(h)}{\partial \dot{\alpha}_t}
& \frac{\partial b(h)}{\partial \sin(\omega t)}
& \frac{\partial b(h)}{\partial \cos(\omega t)}
\end{bmatrix} \\
&=
\begin{bmatrix}
1
& 0
& \sin(\omega t)
& 0
& \alpha_t
& 0
\end{bmatrix}
\end{align*}
With this approximation to $B$, the standard Kalman filter equations given in Section \ref{sec:theKalmanFilter} can be used. The proposed mean reversion approach is thus directly applicable.

For the nonlinear model, the datasets are assumed to approach a fixed mean offset and a fixed mean seasonal amplitude. The attractor distribution thus approximates the central limit of $\gamma_t$ as well as $\alpha_t$. The emission matrix for the attractor distribution is given by
\begin{align*}
B =
\begin{bmatrix}
1 & 0 & 0 & 0 & 0 & 0 \\
0 & 0 & 1 & 0 & 0 & 0
\end{bmatrix}.
\end{align*}

With a two-dimensional attractor distribution, the variance $\Sigma^v_t$ is a two-dimensional matrix. This matrix is configured for an isotropic Gaussian with elements along the diagonal. These elements are manually chosen according to the uncertainty bounds and the slowly varying irregular component of the data.

%% file: results.tex

\section{State-Space Models Results}
\label{sec:results}

\subsection{Methodology}

The datasets are resampled to three samples per day according to \citep{dabrowski2018State}. Resampling simulates handheld sensor readings taken by farmers, where samples are extracted at 05h00, 12h00, and 20h30. Although only 3 of the 96 samples per day are available, the sample rate in the models remains at 96 samples. The remaining 93 samples are treated as missing values that are estimated through filtering and smoothing in the state space models. Forecasts are performed and evaluated over all 96 samples per day.

The time series dataset is split into a training and test set. Filtering is performed on the training set. The attractor distributions are obtained from these filtered results. Forecasts are evaluated on the test set. The location of the split between the training and test sets is specifically chosen around some form of inflection point. At these inflection points, a model without mean reversion is more likely to deviate from the global trend. 

The forecasts are made over multiple steps to provide long-term forecasts. The number of samples over which the forecasts are made are provided in Table \ref{table:forecastSteps}.
\begin{table}[!t]
	\centering
	\footnotesize
	\begin{tabular}{c c c c}
		\toprule
		Dataset & Samples & Time & Frequency  \\
		\midrule
		DO          & 1200 & 12.5 days & 15 min \\
		pH          & 1000 & 10.4 days & 15 min \\
		Temperature & 1100 & 11.5 days & 15 min \\
		\bottomrule
	\end{tabular}
	\caption{Forecast horizon in number of samples as well as time for the datasets used in this demonstration. The last column provides the sample rate of the sensor used to gather the dataset. Forecast horizons are determined by the selected inflection point in the data.}
	\label{table:forecastSteps}
\end{table}

The normalised root mean squared error is used to provide an evaluation of the error between the forecast result and the measured data. Let $\hat{y}_t$ denote the forecast and let $y_t$ denote the true value of some time series at time $t$. For a forecast over $N$ samples, the normalised root mean squared error (NRMSE) is given by
\begin{align}
\epsilon_\text{nrmse} = \frac{ \sqrt{\frac{1}{N} \sum_{i=1}^{N} (y_i - \hat{y}_i)^2}}
{y_\text{max} - y_\text{min}}
\times 100\%
\label{eq_nrmse}
\end{align}
where $y_\text{max}$ and $y_\text{min}$ are the maximum and minimum dataset values respectively. The NMSE for a single sample $i$ is given by
\begin{align}
\epsilon_\text{nrmse} = \frac{ \sqrt{(y_i - \hat{y}_i)^2}}
{y_\text{max} - y_\text{min}}
\times 100\%
\label{eq_sample_nrmse}
\end{align}

\subsection{Linear Model Results}

\begin{figure*}[!t]
	\centering
	\begin{subfigure}[t]{3.06in}
		\includegraphics{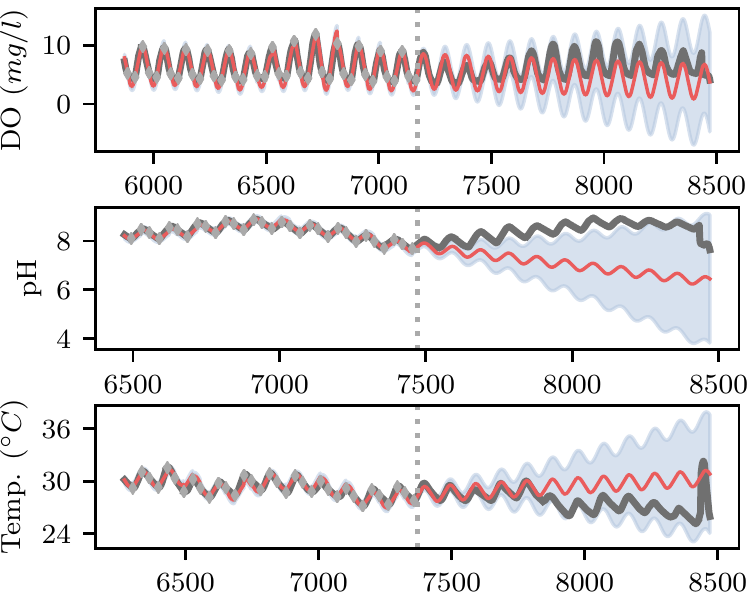}
		\caption{Linear model forecasts without mean reversion}
		\label{fig:kf_forecast_a}
	\end{subfigure} \hfill
	\begin{subfigure}[t]{3.06in}
		\includegraphics[width=3in]{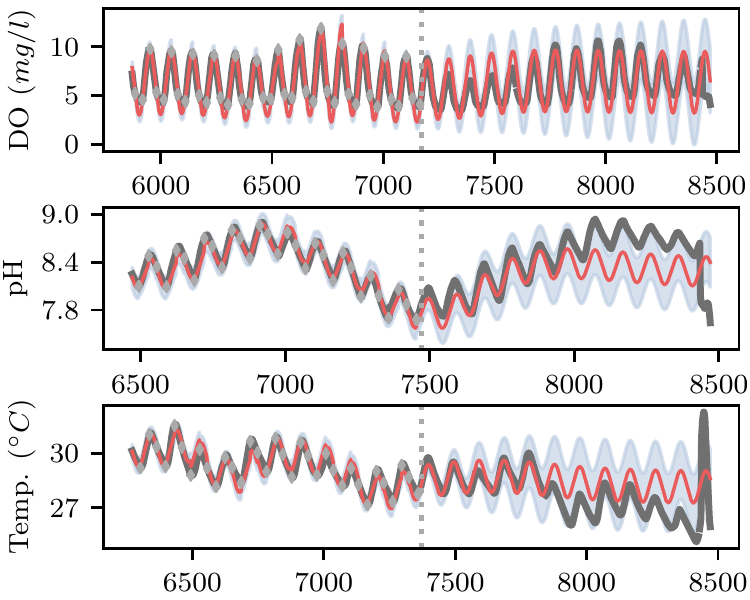}
		\caption{Linear model forecasts with mean reversion}
		\label{fig:kf_forecast_b}
	\end{subfigure}
	\caption{Linear model forecasts of the dissolved oxygen ($mg/l$), pH, and temperature ($^\circ C$) over sample indexes. The red line is a plot of the forecast and the blue filled region is a plot of the forecast standard deviation. The dark grey line is a plot of the sensor data sampled at 15 minute intervals, and the light grey markers indicate sub-samples extracted at 05h00, 12h00, and 20h30. The vertical grey dotted line indicates the start of the forecast. Only the last portion of the historical data are shown.}
	\label{fig:kf_forecast}
\end{figure*}

Plots of the forecasts for the linear model are presented in \figurename{~\ref{fig:kf_forecast}}. The horizontal axes describe the sample number. Without mean reversion, the forecast trends deviate from the ground truth as illustrated in \figurename{~\ref{fig:kf_forecast_a}}. These deviations are due the inflection point in the long-term trend from which the forecasts extend. Reasonable forecasts are obtained up to the end of the first seasonal cycle where variations in the true trend are minimal. After the first cycle, the forecasts begin to deviate as the true trend changes in a non-linear or stochastic manner.

As indicated in \figurename{~\ref{fig:kf_forecast_b}}, enforcing mean reversion provides significant improvements to long term forecasts. Mean reversion draws the deviant forecasts back towards the average of the previously observed dynamics.

The blue filled regions plot the standard deviation of the posterior filtered distribution. This represents the uncertainty in the forecast. As expected, the mean reversion reduces magnitude of the standard deviation through the pseudo observations from the attractor distribution. The level to which the pseudo-observations affect the standard deviation depends on the attractor distribution covariance $\Sigma^v$.

The plots for the pH dataset in \figurename{~\ref{fig:kf_forecast_b}} provide insight into the limitations of the mean reversion approach. The long-term forecasts settle to the attractor distribution mean, while the fluctuations in the trend continue to vary. That is, the slowly-varying fluctuations of the data are not perfectly modelled. These fluctuations are treated as stochastic variations, where there is no deterministic function to model them. Instead, they are modelled by the fixed attractor distribution. Note however that the forecast over the first five days (480 samples) is still accurate and is a significant improvement over the model without mean reversion.
%

\begin{figure*}[!t]
	\centering
	\begin{subfigure}[t]{3.06in}
		\includegraphics{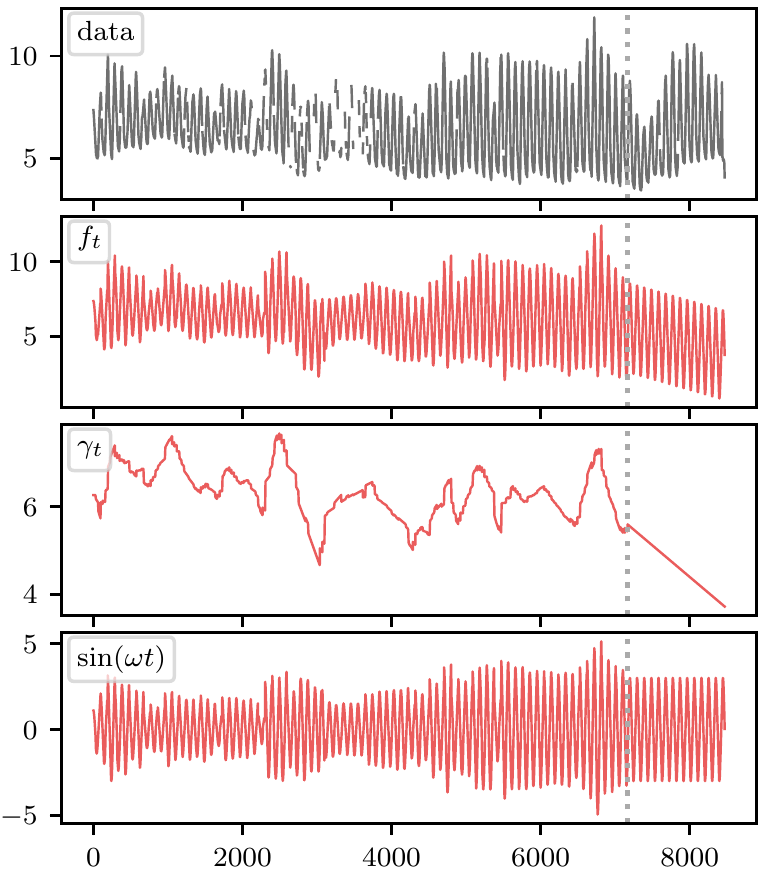}
		\caption{Latent variables for the linear model without mean reversion.}
		\label{fig:kf_hidden_a}
	\end{subfigure} \hfill
	\begin{subfigure}[t]{3.06in}
		\includegraphics{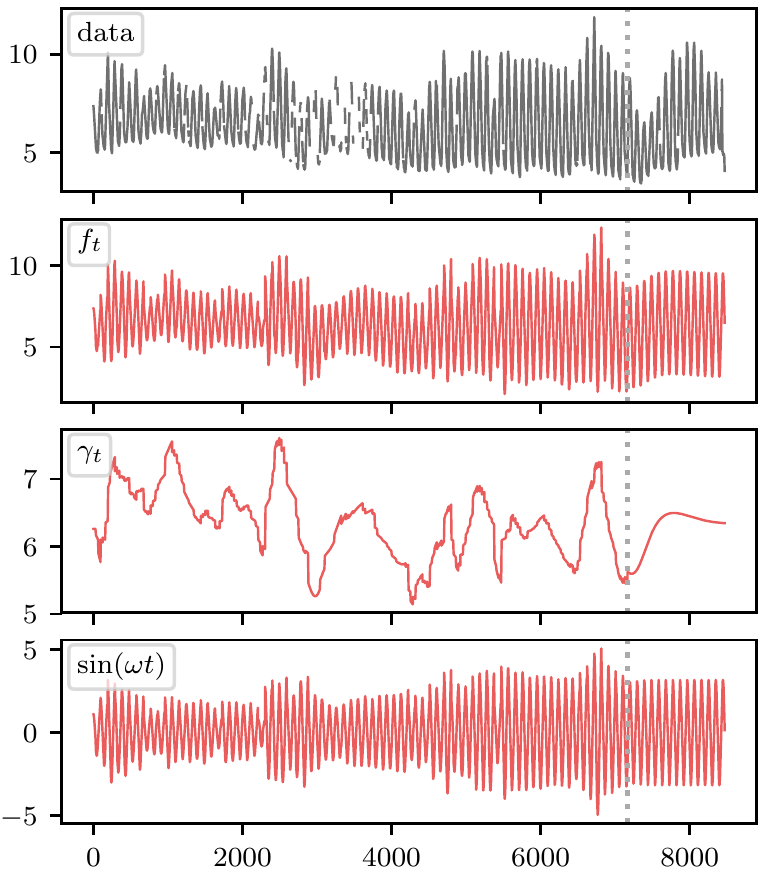}
		\caption{Latent variables for the linear model with mean reversion.}
		\label{fig:kf_hidden_b}
	\end{subfigure}
	\caption{Plots of the data, filtered mean $f_t$, the trend component $\gamma_t$, and the sinusoidal component $\sin(\omega t)$ for the linear model on the dissolved oxygen dataset over the sample index. The gaps in the data plots are due to missing data.}
	\label{fig:kf_hidden}
\end{figure*}

A plot of the linear model's latent variables for the dissolved oxygen dataset is presented in \figurename{~\ref{fig:kf_hidden}}. Mean reversion is applied to the trend component $\gamma_t$. Without mean reversion, the trend of the forecast continues linearly with a steep gradient. Mean reversion causes the trend to curve back towards the attractor distribution mean. By increasing $\Sigma^v$, the time it takes for the curve to settle can be increased. Decreasing $\Sigma^v$ results in a quicker settling time.

Mean reversion is not applied to the sinusoidal component, $\psi_t$. The seasonal oscillation thus continues throughout the forecast. This demonstrates the key feature of the model where mean reversion is applied to one specific component in the model.
%

The NRMSE over the complete forecast for all datasets is presented in Table \ref{table:kf_error}. The results show that mean reversion produces significant improvements in forecast ability. Though the RMSE for the mean reversion in the pH dataset is high, it is a significant improvement over the linear model without mean reversion.
%
\begin{table}[!t]
    \centering
    \footnotesize
    \begin{tabular}{c c c}
        \toprule
        Dataset & Without MR & With MR \\
        \midrule
        DO & 29.69 & 16.68 \\
        pH & 116.69 & 21.90 \\
        Temperature: & 31.09 & 16.20 \\
        \bottomrule
    \end{tabular}
    \caption{NRMSE of the linear model with and without mean reversion (MR) over the entire forecast presented in \figurename{~\ref{fig:ekf_forecast}}.}
    \label{table:kf_error}
\end{table}

\begin{figure*}[!t]
	\centering
	\begin{subfigure}[t]{3.06in}
		\includegraphics{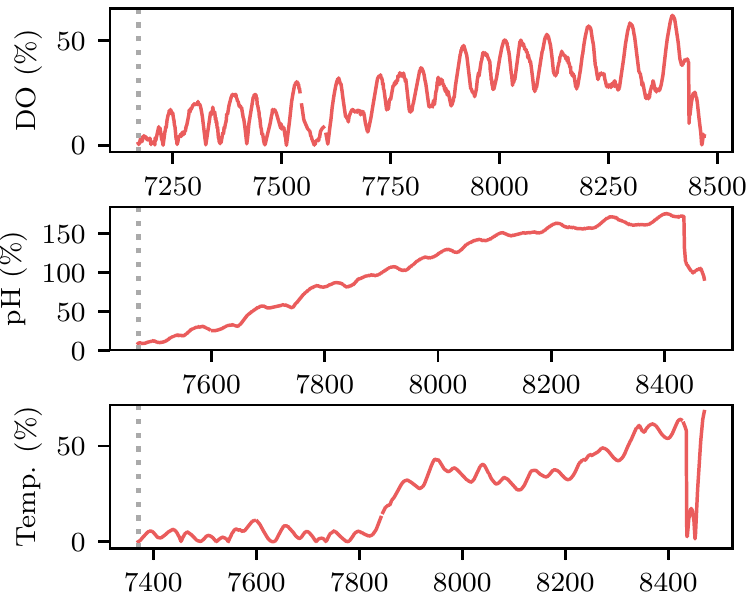}
		\caption{NRMSE for the linear model without mean reversion.}
		\label{fig:kf_error_a}
	\end{subfigure} \hfill
	\begin{subfigure}[t]{3.06in}
		\includegraphics{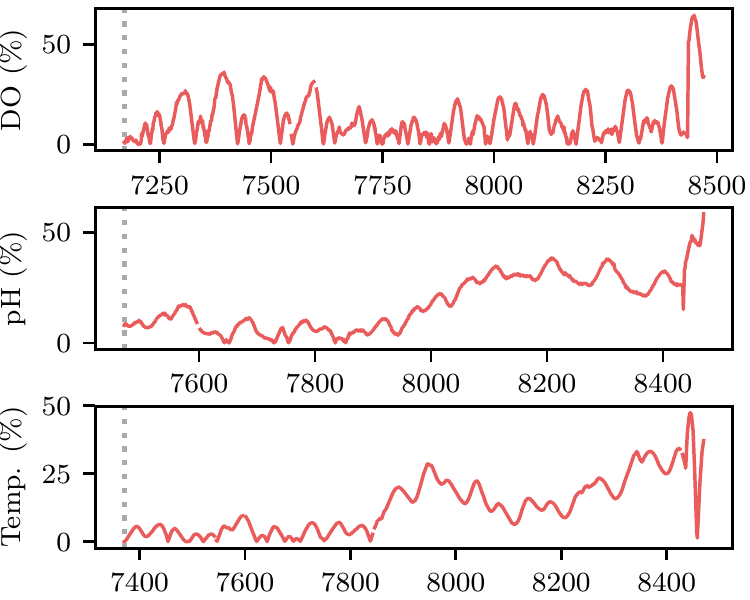}
		\caption{NRMSE for the linear model with mean reversion.}
		\label{fig:kf_error_b}
	\end{subfigure}
	\caption{Per-sample NRMSE (equation (\ref{eq_sample_nrmse})) for the linear model forecasts on the DO dataset presented in \figurename{~\ref{fig:kf_forecast}}.}
	\label{fig:kf_error}
\end{figure*}

A plot of the per-sample NRMSE error (equation (\ref{eq_sample_nrmse})) for the forecast is plotted in \figurename{~\ref{fig:kf_error}}. The error for the model without mean reversion increases over the forecast time. This demonstrates that the forecast deviates from the ground truth with increasing forecast reach. For the model with mean reversion, the error remains relatively constant over the entire forecast. This demonstrates that the model performs equally well at short and long-term forecasting. This is especially remarkable as the model is forecasting more than 1000 steps-ahead in time.
%

\subsection{Nonlinear Model Results}

\begin{figure*}[!t]
	\centering
	\begin{subfigure}[t]{3.06in}
		\includegraphics{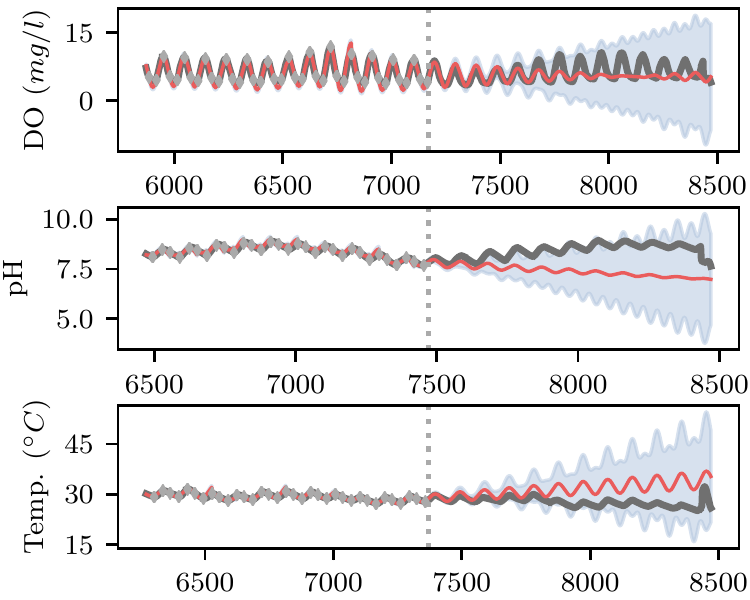}
		\caption{Nonlinear model forecasts without mean reversion}
		\label{fig:ekf_forecast_a}
	\end{subfigure} \hfill
	\begin{subfigure}[t]{3.06in}
		\includegraphics{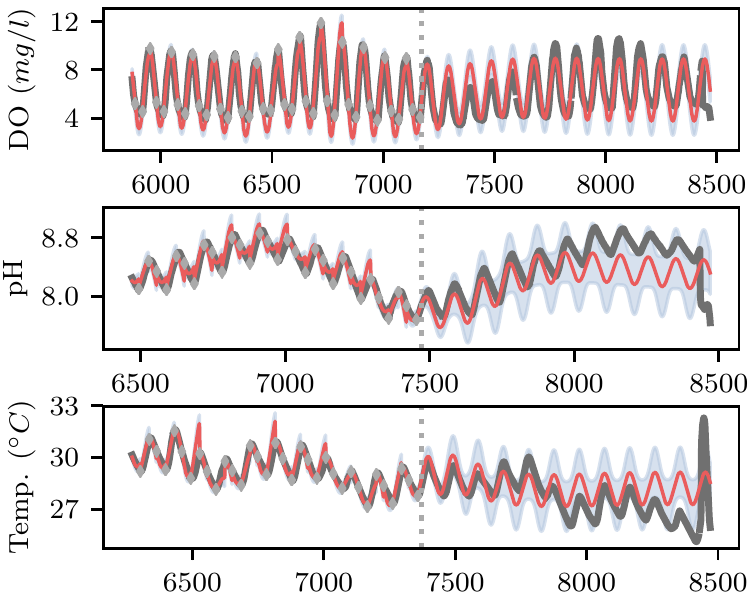}
		\caption{Nonlinear model forecasts with mean reversion}
		\label{fig:ekf_forecast_b}
	\end{subfigure}
    \caption{Nonlinear model forecasts of dissolved oxygen ($mg/l$), pH, and temperature ($^\circ C$) over sample indexes. The red line is a plot of the forecast and the blue filled region is a plot of the forecast standard deviation. The dark grey line is a plot of the sensor data sampled at 15 minute intervals, and the light grey markers indicate sub-samples extracted at 05h00, 12h00, and 20h30. The vertical grey dotted line indicates the start of the forecast. Only the last portion of the historical data are shown.}
	\label{fig:ekf_forecast}
\end{figure*}

Plots of the forecasts for the nonlinear model are presented in \figurename{~\ref{fig:ekf_forecast}}. As for the linear model, mean reversion provides significant improvement in the forecasts and reduces the uncertainty in the forecast.
%

As illustrated in \figurename{~\ref{fig:ekf_forecast_a}}, the oscillation component decays over the forecast of the DO dataset. This follows the trend in the data leading up to the forecast, where the oscillation amplitude is decreasing. The trend in the data however does not continue decreasing as it does in the forecast. Mean reversion is thus applied to both the trend component $\psi_t$ and the amplitude component $\alpha_t$. The result is that both of these components are corrected to provide a more accurate forecast.

\begin{figure*}[!t]
	\centering
	\begin{subfigure}[t]{3.06in}
		\includegraphics{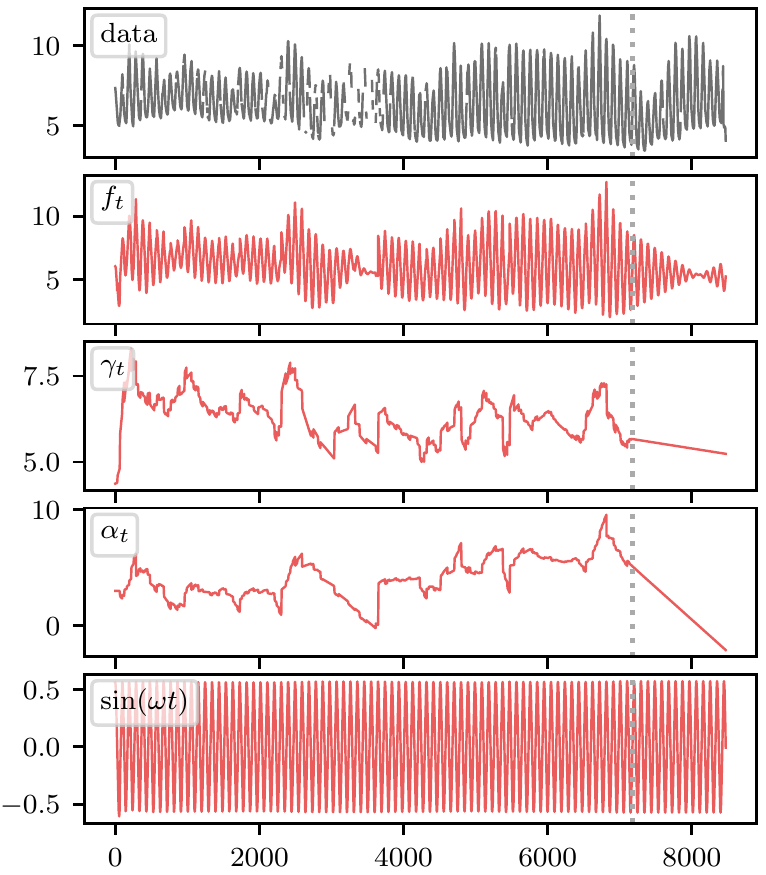}
		\caption{Latent variables for the nonlinear model without mean reversion.}
		\label{fig:ekf_hidden_a}
	\end{subfigure} \hfill
	\begin{subfigure}[t]{3.06in}
		\includegraphics{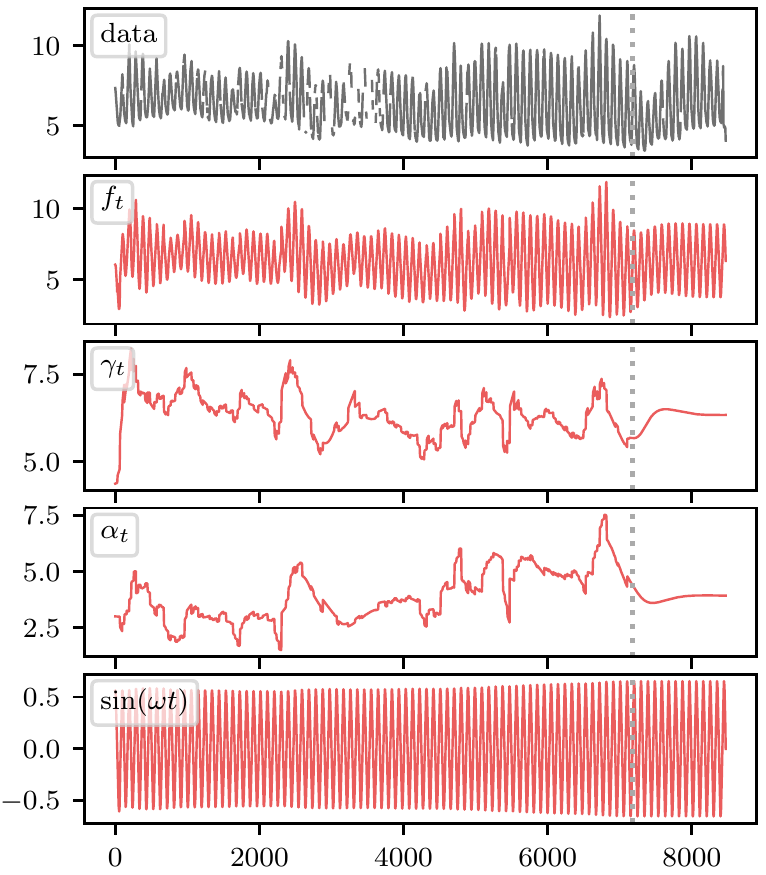}
		\caption{Latent variables for the nonlinear model with mean reversion.}
		\label{fig:ekf_hidden_b}
	\end{subfigure}
	\caption{Plots of the data, filtered mean $f_t$, the trend component $\gamma_t$, the sinusoidal component $\sin(\omega t)$, and the amplitude component $\alpha_t$ for the nonlinear model over the sample index. The gaps in the data plots are due to missing data.}
	\label{fig:ekf_hidden}
\end{figure*}

A plot of the latent variables for the DO dataset are presented in \figurename{~\ref{fig:ekf_hidden}}. The amplitude of the $\sin(\omega t)$ component remains fairly constant when compared to the linear model. This is expected as $\alpha_t$ and $\sin(\omega t)$ are separated in the nonlinear model, whereas in the linear model, they are combined into a single component. Both the trend $\gamma_t$ and amplitude $\alpha_t$ components are affected by the inflection point in the data where the forecast begins. They both veer off with a steep gradient. Mean reversion is applied to correct $\gamma_t$ and $\alpha_t$, and draw them back to the mean. The seasonal component is left to oscillate throughout the forecast.
%

The NRMSE over the entire forecast for all datasets is presented in Table \ref{table:ekf_error}. As for the linear model, the mean reversion reduces the error. Comparing the linear model results in Table \ref{table:kf_error} and the nonlinear model results in Table \ref{table:ekf_error}, it is clear that the nonlinear model achieves the best results. The nonlinear model is however a more complex model.
%
\begin{table}[!t]
    \centering
    \footnotesize
    \begin{tabular}{c c c}
        \toprule
        Dataset & Without MR & With MR \\
        \midrule
        DO & 25.12 & 14.44 \\
        pH & 87.89 & 21.84 \\
        Temperature: & 64.48 & 16.15 \\
        \bottomrule
    \end{tabular}
    \caption{NRMSE of the nonlinear model with and without mean reversion (MR) over the forecast presented in \figurename{~\ref{fig:ekf_forecast}}.}
    \label{table:ekf_error}
\end{table}

\begin{figure*}[!t]
	\centering
	\begin{subfigure}[t]{3.06in}
		\includegraphics{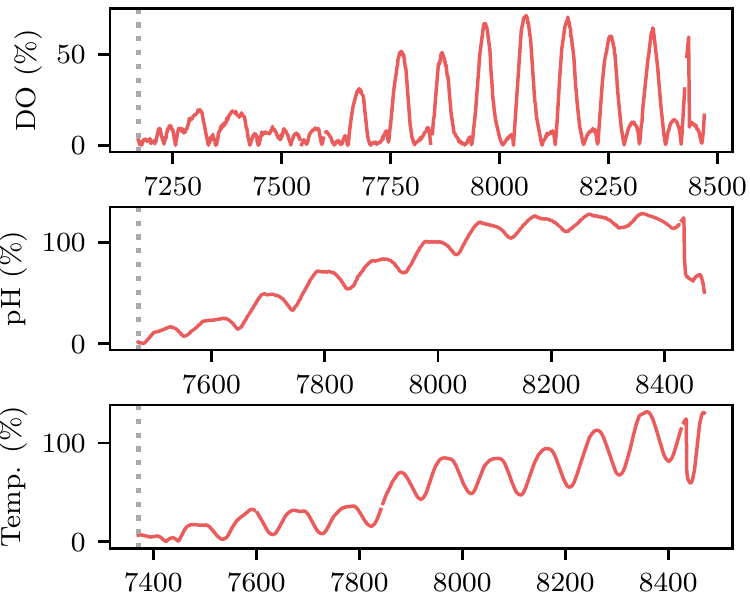}
		\caption{NRMSE for the nonlinear model without mean reversion.}
		\label{fig:ekf_error_a}
	\end{subfigure} \hfill
	\begin{subfigure}[t]{3.06in}
		\includegraphics{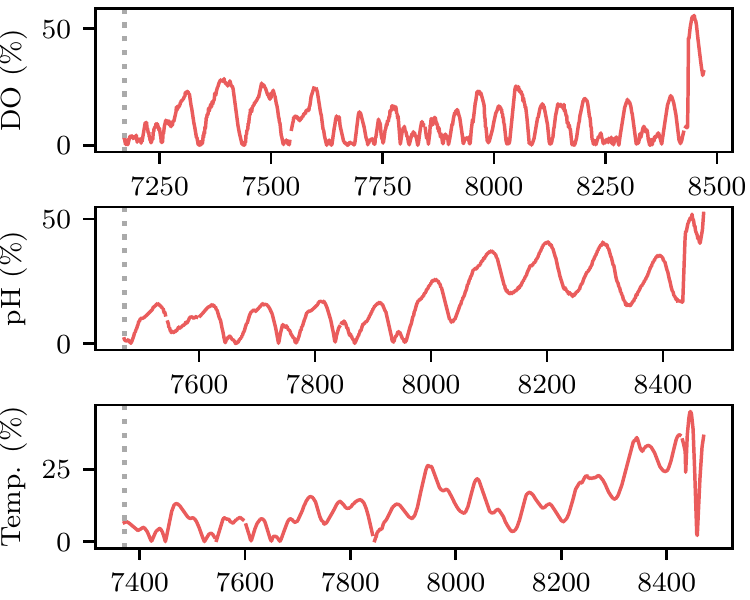}
		\caption{NRMSE for the nonlinear model with mean reversion.}
		\label{fig:ekf_error_b}
	\end{subfigure}
	\caption{Per-sample NRMSE (equation (\ref{eq_sample_nrmse})) for the nonlinear model forecasts on the DO dataset presented in \figurename{~\ref{fig:ekf_forecast}}.}
	\label{fig:ekf_error}
\end{figure*}

A plot of the per-sample NRMSE error (equation (\ref{eq_sample_nrmse})) is presented in \figurename{~\ref{fig:ekf_error}}. As for the linear model, mean reversion reduces the error in the long-term forecasts.
%
%

%% file: comparisonResults.tex

\section{Time Series Model Comparison}
\label{sec:comparison}

A comparison between a LDS \citep{dabrowski2018State}, a dynamic linear model (DLM) \citep{west2013bayesian}, a seasonal autoregressive integrated moving average (SARIMA) model, and Facebook's Prophet model \citep{Taylor2018Forecasting} is performed.

The linear LDS model of \citep{dabrowski2018State} is used as described in section \ref{sec:linearModel}. The DLM model is a free-form seasonal model \citep{west2013bayesian} with a first order trend component as used in the LDS. Mean reversion using equation (\ref{eq_mr}) and weighted mean reversion using equation (\ref{eq_wmr}) is applied to the trend components in the LDS and DLM models. The weighted mean reversion is applied with $\lambda = 0.1$. In tables and figures, models using mean reversion and weighted mean reversion are denoted by a `MR' and a `WMR' subscript respectively.

The SARIMA(5,1,3)(0,1,0)96 model\footnote{https://www.statsmodels.org} is used on all datasets. The model order was chosen according to autocorrelation and partial autocorrelation plots. The Prophet model\footnote{https://facebook.github.io/prophet/} is configured with a linear growth trend, an additive daily seasonal component, and an interval width of 0.8.

The set of models are compared on the dissolved oxygen, pH, and temperature datasets. In this comparison, the datasets are not resampled as was done in section \ref{sec:results}. All 96 samples per day are used in all models. Each model provides a 10 day (960 sample) forecast from the set of 10 pre-selected random starting points. Ten days is selected as it represents a reasonable long-term forecast in this application. The average NRMSE over the 10 forecasts for each model and dataset are presented in Table \ref{table:comparisonResults}.
%
\begin{table*}[!t]
	\centering
	\footnotesize
	\begin{tabular}{c c c c c c c c c}
		\toprule
		Dataset & LDS & $\text{LDS}_\text{MR}$ & $\text{LDS}_\text{WMR}$ & DLM & $\text{DLM}_\text{MR}$ & $\text{DLM}_\text{WMR}$ & SARIMA & Prophet  \\
		\midrule
		DO           & 33.51 & 14.41 & 15.98 & 25.41 & \textbf{10.81} & 11.08 & 15.27 & 16.07 \\
		pH           & 60.14 & 35.03 & 27.61 & 61.74 & 34.36 & \textbf{24.76} & 65.13 & 25.61 \\
		Temperature  & 107.92 & 38.14 & 34.33 & 104.80 & 36.83 & \textbf{31.86} & 109.26 & 71.12 \\
		Average      & 67.19 & 29.19 & 25.97 & 63.98 & 27.33 & \textbf{22.56} & 63.22 & 37.60 \\
		\bottomrule
	\end{tabular}
	\caption{Average NRMSE error (\%) over ten 960-step-ahead forecasts for the set of models and datasets. Mean reversion is denoted by MR. Weighted mean reversion is denoted by WMR.}
	\label{table:comparisonResults}
\end{table*}

The LDS performs poorly over a long-term forecast. However, when using the mean reversion, the forecast is significantly improved. Using weighted mean reversion provides further improvements on the pH and temperature datasets.

The DLM generally does better than the LDS. It is a more complex model and is able to provide a more refined representation of the seasonal curves. This increased complexity comes at a significant cost with a 97-dimensional state vector. This can be problematic in hardware where computational power and memory are limited. In comparison with the DLM, the LDS has a 4-dimensional state vector. The LDS thus performs surprisingly well in comparison.

The DLM with weighted mean reversion provides the lowest average NRMSE results over all datasets. Other than the pH dataset, the other mean reversion model variants take the second, third and fourth place. For the pH dataset, the Prophet model provides highly competitive results and takes second place. The SARIMA model performs well on the dissolved oxygen dataset, otherwise it provides similar results to the DLM and LDS models.

The SARIMA model has first order differencing and the Prophet model has a linear growth trend. These components function as linear trend components. Thus, like state-space models, the SARIMA and Prophet models are susceptible to forecast deviations. Given this, the Prophet model performs remarkably well.

\begin{figure}[!t]
	\centering
	\includegraphics{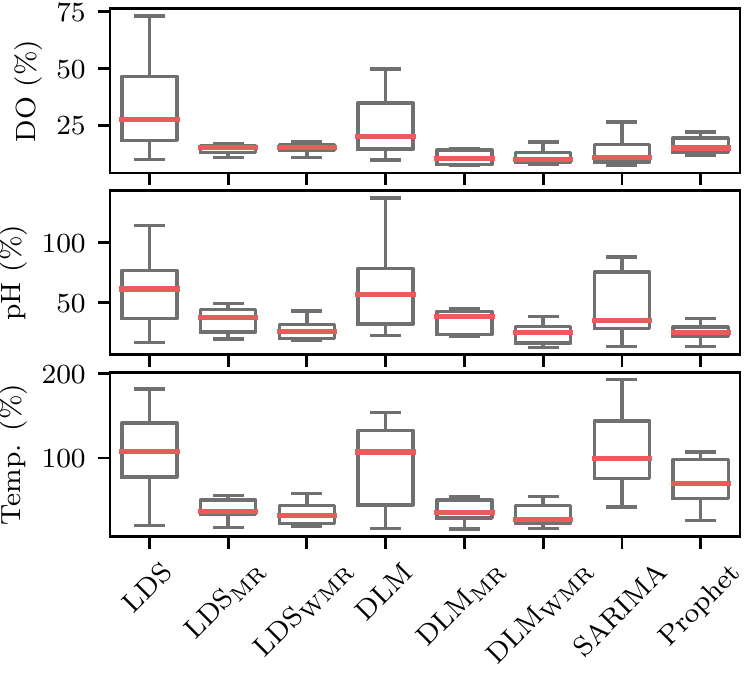}
	\caption{Box-whisker plots comparing the set of models over each dataset for the NRMSE results.}
	\label{fig:boxplots}
\end{figure}

To illustrate the robustness of the models and the statistical significance of the results, box-whisker plots are presented in \figurename{~\ref{fig:boxplots}}. In the absence of mean reversion, the LDS and DLM models produce results with high NRMSE values and large boxes. The large boxes indicate a high variation in the forecast accuracy. Introducing mean reversion or weighted mean reversion both increases accuracy and reduces variation in the forecasts. The result is a more robust model.

For the pH and temperature datasets, the $\text{DLM}_\text{WMR}$ model produces boxes which are below the LDS, DLM, SARIMA, and Prophet model boxes. This indicates some level of statistical significance that the $\text{DLM}_\text{WMR}$ outperforms these models.
%

The computation times are presented in Table \ref{table:processingTimes}. These times include the parameter estimation as well as the forecasting operations. All models are implemented in Python and run on a Dual-Core Intel i5 processor. The mean reversion increases the processing time as the pseudo samples are required to be calculated. Weighted mean reversion further increases computational complexity resulting in further increased processing times. Weighted mean reversion in the LDS is still however quicker than the Prophet and SARIMA models. The SARIMA model has the highest processing time, which is primarily due to the parameter estimation operation. Compared with the DLM, the Prophet model is more computationally efficient.

\begin{table*}[!t]
    \centering
    \footnotesize
    \begin{tabular}{c c c c c c c c c}
        \toprule
        Dataset & LDS & $\text{LDS}_\text{MR}$ & $\text{LDS}_\text{WMR}$ & DLM & $\text{DLM}_\text{MR}$ & $\text{DLM}_\text{WMR}$ & SARIMA & Prophet  \\
        \midrule
        DO & 1.96 & 3.61 & 5.12 & 31.83 & 64.38 & 97.17 & 556.43 & 15.22 \\
        pH & 1.85 & 3.47 & 5.18 & 32.1 & 64.38 & 97.22 & 87.85 & 17.73 \\
        Temperature & 2.16 & 3.95 & 5.61 & 31.6 & 64.53 & 96.46 & 190.23 & 17.51 \\
        \bottomrule
    \end{tabular}
    \caption{Average processing time in seconds  over ten 960-step-ahead forecasts for the set of models and datasets. Mean reversion is denoted by MR. Weighted mean reversion is denoted by WMR.}
    \label{table:processingTimes}
\end{table*}
%

%% file: conclusion.tex

\section{Summary and Conclusion}
\label{sec:conclusion}

In this study a novel mean reversion approach is presented for state-space models. The mean reversion is performed using an attractor distribution with a Gaussian form. The mean of this distribution is approximated by the average filtered estimate over previously observed samples. This mean provides an approximation of the average dynamics over the sequence. To draw a forecast towards the mean, filtering is applied with pseudo-observations obtained from attractor distribution. The result is that the forecast converges to the attractor distribution mean in the limit.

We demonstrate the approach with a linear and nonlinear LDS in a prawn pond water quality forecasting application. Results show a significant improvement in long-term forecasts. Furthermore, a comparison between various time series models on the prawn pond water quality dataset is presented. The results demonstrate that the lowest errors are obtained when weighted mean reversion is used in the DLM.

A limitation of the attractor distribution is that it is stationary. The result is that the long-term forecast is drawn to a fixed mean. In future work, a non-stationary attractor distribution could be investigated. The result would be that the forecast would be drawn to a particular dynamic rather than a fixed mean. Future work could also include an investigation into estimating the attractor distribution covariance matrix $\Sigma^v_t$ using the expectation maximisation algorithm. 

Finally, though the proposed approach is demonstrated on an aquaculture problem, it is applicable to other problems with similar properties. Future work could include testing the approach on problems such as weather-related forecasting, electricity load forecasting, algal bloom forecasting, and other environmental applications with seasonal data.